\documentclass[sigconf]{acmart}
\usepackage{graphicx}
\usepackage{xcolor}
\usepackage{multirow}
\usepackage{enumitem}
\usepackage{pifont}
\usepackage{makecell}
\usepackage{float}

\usepackage{seqsplit} 
\renewcommand{\seqinsert}{\ifmmode\allowbreak\else\-\fi}

\hypersetup{
    colorlinks=true,
    linkcolor=black,
    citecolor=black,
    urlcolor=black
}

\newcommand{\insightbox}[1]{%
  \noindent\fbox{%
    \begin{minipage}{0.46\textwidth}
      #1
    \end{minipage}
  }%
}

\newcommand{\revised}[1]{{\leavevmode\color{black}{#1}}}

\begin{document}
\begin{sloppypar}


\title{\revised{Towards Understanding and Characterizing Vulnerabilities in Intelligent Connected Vehicles through Real-World Exploits}}

\author{Yuelin Wang$^{1}$, 
Yuqiao Ning$^{2}$, 
Yanbang Sun$^{1}$, 
Xiaofei Xie$^{3}$, 
Zhihua Xie$^{1}$, 
Yang Chen$^{2}$, 
Zhen Guo$^{2}$, 
Shihao Xue$^{2}$, 
Junjie Wang$^{1*}$,
Sen Chen$^{4}$\\
$^{1}$College of Intelligence and Computing, Tianjin University, China\\
$^{2}$China Automobile Data of Tianjin Co., Ltd., China Automotive Technology \& Research Center Co., Ltd., China\\
$^{3}$Singapore Management University, Singapore\\
$^{4}$Nankai University, China}
\thanks{\textsuperscript{*}Junjie Wang is the corresponding author.}

\renewcommand{\shortauthors}{Wang et al.}





\begin{abstract}
Intelligent Connected Vehicles (ICVs) are a core component of modern transportation systems, and their security is crucial as it directly relates to user safety.
\revised{Despite prior research, most existing studies focus only on specific sub-components of ICVs due to their inherent complexity. As a result, there is a lack of systematic understanding of ICV vulnerabilities. Moreover, much of the current literature relies on human subjective analysis, such as surveys and interviews, which tends to be high-level and unvalidated, leaving a significant gap between theoretical findings and real-world attacks.}

To address this issue, we conducted the first large-scale empirical study on ICV vulnerabilities. 
\revised{We began by analyzing existing ICV security literature and summarizing the prevailing taxonomies in terms of vulnerability locations and types. To evaluate their real-world relevance,} we collected a total of 649 exploitable vulnerabilities, including 592 from eight ICV vulnerability discovery competitions, Anonymous Cup, between January 2023 and April 2024, covering 48 different vehicles.
The remaining 57 vulnerabilities were submitted daily by researchers.
\revised{Based on this dataset, we assessed the coverage of existing taxonomies and identified several gaps, discovering one new vulnerability location and 13 new vulnerability types.}
We further categorized these vulnerabilities into 6 threat types (e.g., privacy data breach) and 4 risk levels (ranging from low to critical) and analyzed participants’ skills and the types of ICVs involved in the competitions.
This study provides a comprehensive and data-driven analysis of ICV vulnerabilities, offering actionable insights for researchers, industry practitioners, and policymakers.
To support future research, we have made our vulnerability dataset publicly available.

\end{abstract}

\maketitle

\section{Introduction}\label{sec: intro}
With the rapid advancement of artificial intelligence and 5G communication technologies, Intelligent Connected Vehicles (ICVs) are becoming integral to modern transportation systems~\cite{gan2024large}.
Unlike traditional vehicles, ICVs incorporate cloud platforms, mobile applications (APPs), In-Vehicle Infotainment (IVI) systems, Advanced Driver Assistance Systems (ADAS), and diverse networks, markedly enhancing their intelligence~\cite{ICVtest2023}.
However, this complex multi-component architecture also introduces significant security risks, which may lead to serious consequences.
For example, in 2018, a self-driving Uber vehicle fatally struck a pedestrian due to a failure in the ADAS to correctly identify her~\cite{Pedestrian2018}.
In September 2024, Rivera et al.~\cite{Kia} discovered an authorization vulnerability in Kia ICVs, allowing attackers to remotely control the vehicle using only the license plate number, posing a serious threat to user safety.

\revised{
To address the challenges in ICV security and safety, researchers have conducted extensive explorations in ICV security and proposed several theoretical attack taxonomies and threat models~\cite{pekaric2021taxonomy, de2024security, gul2024vehicle, pham2021survey, sommer2019survey}.
For example, Pekaric et al.~\cite{pekaric2021taxonomy} systematically summarized existing work, identified 48 attack vectors, and categorized them into five types (e.g., physical attacks), providing a theoretical foundation for developing defense strategies.
Moreover, Jing et al.~\cite{jing2024revisiting} conducted surveys and interviews with automotive security practitioners and systematically categorized the attack surfaces of ICVs into seven major classes (e.g., in-vehicle components), constructing a structured attack surface model for ICVs.
These early taxonomies offer valuable frameworks for understanding ICV security.

However, existing taxonomies suffer from two main limitations. First, the attacks and analyses they are based on are often incomplete and focus only on specific aspects of ICVs, such as physical and network-based attacks~\cite{pekaric2021taxonomy}, or issues in autonomous driving systems like Apollo~\cite{9284001,tang2023survey}. As a result, there is a lack of holistic understanding of the overall security landscape in ICVs.
Second, most taxonomies are constructed based on expert knowledge and literature review, resulting in high-level, conceptual abstractions of the attack surface. These taxonomies remain largely theoretical and have not been systematically validated against real-world attack data.
In particular, it remains unclear which categories in existing taxonomies are actually reflected in real-world attacks, and which potential categories are missing altogether.
Addressing this gap is crucial for evaluating the practical coverage of current taxonomies, and for identifying areas that require refinement or extension.
}

\revised{To this end, we conducted the first large-scale empirical study ICV vulnerabilities.
To address the limitations of existing research, we first collected and analyzed 13 representative papers on ICV security, manually extracting the reported attack methods, vulnerabilities, and security issues.
Based on this, we constructed a \textit{unified taxonomy} focused on two key dimensions: vulnerability locations and vulnerability types, encompassing 11 location categories and 35 type categories.
This unified taxonomy provides a comprehensive synthesis of existing ICV vulnerability analyses in the literature.}

To evaluate the real-world applicability of these taxonomies, we made significant efforts to collect and analyze real-world exploits.
Over a 16-month period, we gathered 890 vulnerability reports through daily submissions from researchers and eight ``Anonymous Cup'' ICV vulnerability discovery competitions. These competitions employed a black-box testing approach on complete vehicles and involved 48 vehicle models \revised{from 19 manufacturers}, encompassing a wide range of vehicle architectures (e.g., distributed ECUs) and security mechanisms (e.g., QNX hypervisor).

All submitted reports underwent a rigorous validation process—including real-vehicle reproduction, expert review, and deduplication, which ultimately confirmed 649 unique and exploitable vulnerabilities.
The entire data collection and analysis effort spanned \revised{98} person-months, with vehicle rental costs totaling USD 51,000 and vulnerability bounties exceeding USD 340,000.
These efforts ensured the scale, diversity, and reliability of the dataset, forming a robust foundation for validating and extending existing ICV vulnerability taxonomies.


\revised{
Based on the collected taxonomies and real-world exploitable vulnerabilities, we systematically mapped the 649 real-world vulnerabilities to these taxonomies, which allows us to evaluate how well these empirical vulnerabilities align with existing security classification taxonomies.
Our analysis shows that the collected vulnerabilities could cover 91\% of the subcategories in location-based taxonomies and 89\% of the subcategories in attack-type taxonomies, indicating strong representativeness and broad applicability. 
Moreover, we identified several critical dimensions that frequently appear in real-world attacks but are missing from current taxonomies, including one previously undefined location (i.e., T-Box) and thirteen new attack types (e.g., file accessing vulnerabilities).
These findings not only reveal blind spots in existing vulnerability classification systems but also provide clear, data-driven directions for their future updates, thereby helping security research better align with the actual threat landscape of modern ICV systems.
}

Further analysis results show that vulnerabilities are primarily concentrated in the cloud platform (37.8\%) and IVI module (32.0\%).
In terms of vulnerability types, authorization vulnerabilities (36.4\%) and information leakage (17.3\%) vulnerabilities are the most common, posing serious threats to user privacy and vehicle control.
We also observed strong correlations between vulnerability types and modules. 
For example, all 26 web-related vulnerabilities were found in the cloud platform.
\revised{More importantly, we identified significant differences in vulnerability distribution across vehicle types.
For instance, SUV models have a significantly higher number of vulnerabilities compared to sedans and MPVs, with vulnerabilities particularly concentrated in the IVI module.
These findings reveal the specific security risks faced by different modules and vehicle types, highlighting the importance of developing modular and differentiated security protection strategies.}
Based on these findings, we offer targeted recommendations to ICV stakeholders, including manufacturers, researchers, and regulators. By implementing these recommendations, stakeholders can better mitigate security risks, improve protection across different ICV modules, and ultimately contribute to a safer intelligent transportation ecosystem.


In this paper, we make the following contributions:
\begin{itemize}[leftmargin=*]
    \item 
    \textbf{Large-Scale ICV vulnerability dataset construction.}
    We amassed a dataset of 649 verified ICV vulnerabilities through eight competitions held from January 2023 to April 2024, validated individually with participants to ensure accuracy. This realistic annotated dataset, which is significantly larger than those used in prior studies, is publicly available~\cite{dataset} and provides a robust empirical foundation for future ICV security research.
    \item
    \revised{\textbf{Summarizing, mapping, and extending existing taxonomies.}
We summarized and unified the vulnerability taxonomies derived from existing literature. We mapped the collected vulnerabilities to these taxonomies to evaluate their coverage and relevance. Through this process, we identified and added previously unaddressed vulnerability types and locations, thereby extending the existing classification systems.}

    \item 
    \textbf{Systematic multi-dimensional vulnerability analysis.}
    We conducted an in-depth analysis of vulnerability characteristics from multiple dimensions, uncovering various distribution patterns and inter-dimensional correlations, thereby providing more precise and practical guidance for ICV development and testing.

\end{itemize}
\section{Background}

To deepen the understanding of ICV security, this section introduces the overall architecture of ICVs and related work.

\begin{figure}[t]
\centering
\includegraphics[width=0.88\linewidth]{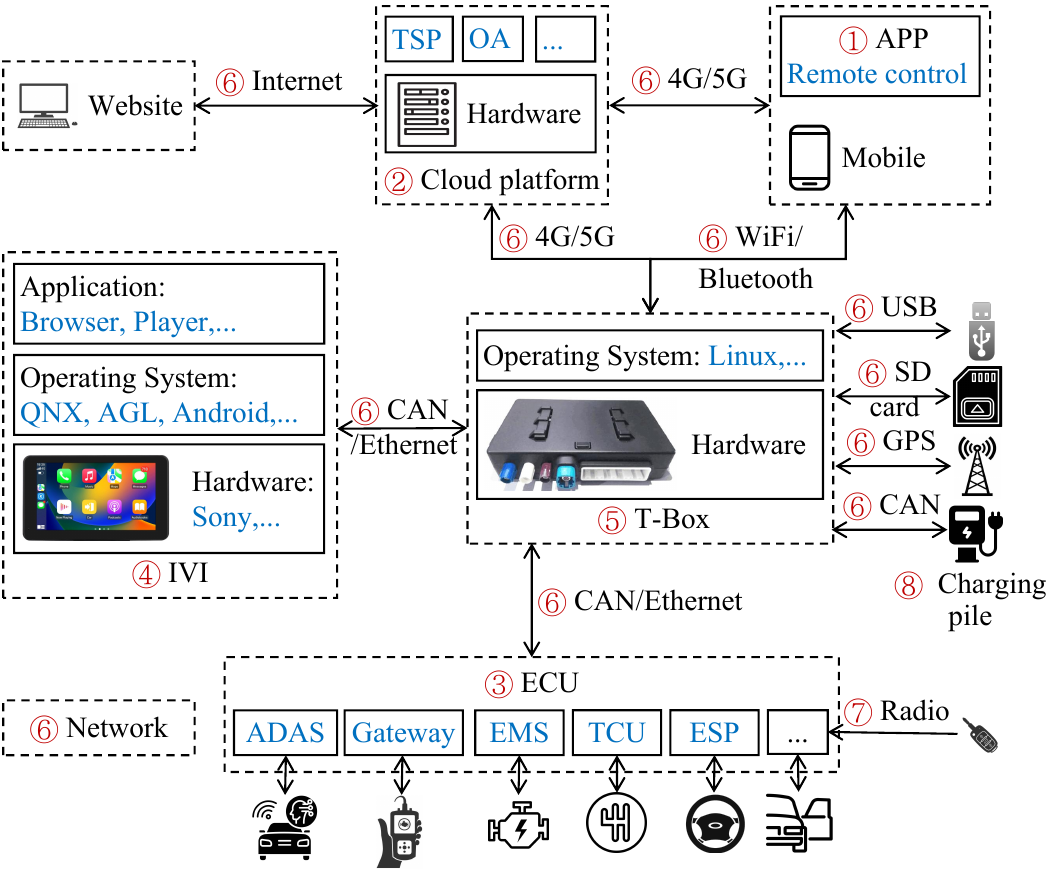}
\captionsetup{justification=centering}
\vspace{-4mm}
\caption{The general architecture of an ICV.}
\label{fig:architecture}
\vspace{-4mm}
\end{figure}

\subsection{ICV architecture}\label{sec: icv archi}

As shown in Fig~\ref{fig:architecture}, the ICV system comprises multiple key components that collectively support its intelligent and connected functionalities.
Among them, the APP (\ding{182}) serves as a vital interface between users and vehicles.
Through APPs developed by ICV enterprises, owners can remotely control their vehicles, enhancing the overall user experience.
These APPs have become a core part of the ICV software system.
Acting as the central hub, the cloud platform (\ding{183}) not only enables services such as remote control, ADAS, OTA updates, and accident assistance, but also provides enterprises with capabilities for data management and productivity improvement.
Another essential component is the ECU (\ding{184}), which governs the operation of various vehicle subsystems.
A typical ICV is equipped with hundreds of ECUs to ensure coordinated system performance.
The IVI system (\ding{185}) delivers both information and entertainment services, supports interactions via touchscreen and smart voice assistants, and runs operating systems such as QNX or Android.

In addition, the T-Box module (\ding{186}) manages communication between internal and external components of the ICV—connecting with the cloud platform via 4G/5G, linking to mobile devices through Wi-Fi or Bluetooth, and interfacing with ECUs via the CAN bus.
ICVs rely on a variety of network technologies (\ding{187}) to perform data transmission and positioning tasks.
These include GPS for positioning, Wi-Fi and Bluetooth for short-range communication, and CAN bus and Ethernet for high-bandwidth in-vehicle data exchange.
Radio components (\ding{188}), such as contactless keys and GPS receivers, are primarily used for vehicle unlocking and positioning.
Finally, as essential infrastructure for electric ICVs, charging piles (\ding{189}) support user authentication via RFID cards, APPs, or QR codes, and, after verification by the cloud platform, negotiate charging parameters with the vehicle using protocols such as ISO 15118 to ensure a safe and efficient charging process.

\subsection{Related work}
\vspace{-5mm}
\revised{
\subsubsection{{Autonomous driving security and safety}}
\label{sec:adsrelatedwork}
Numerous studies have focused on Autonomous Driving Systems (ADSs), which are specialized subsystems comprising perception, prediction, planning, and control modules designed to enable autonomous vehicle operation through advanced AI models. For example, Cheng et al. proposed BehAVExplor~\cite{10.1145/3597926.3598072}, and Li et al.~\cite{9251068} introduced AV-FUZZER—both aiming to uncover issues in ADSs that could lead to vehicle collisions or response timeouts. These works primarily target the planning module and emphasize functional correctness (e.g., model robustness) rather than typical security vulnerabilities. Tang et al.~\cite{tang2023survey} provided a systematic review of testing approaches for ADSs at both the module and system levels. While their review includes some attack scenarios in sensor and perception modules, these are primarily used to illustrate testing optimization objectives rather than to comprehensively analyze security threats. As a result, many aspects of general ICV security are not covered due to the paper’s distinct focus on testing methodologies for ADSs.

Additionally, Pham et al. conducted an empirical survey~\cite{pham2021survey} on attack and defense strategies in autonomous vehicles, covering threats such as LiDAR/Radar jamming and adversarial attacks on the perception module of ADS. Gao et al.~\cite{9625017} analyzed the attack surface from four dimensions: sensors  (e.g., camera), operating systems (e.g., Apollo), control systems (e.g., CAN), and V2X communication (e.g., V2V, V2I), which are needed for the autonomous driving.  Garcia et al.~\cite{9284001} conducted pioneering work on general bugs in open-source ADSs such as Apollo~\cite{Apollo} and Autoware~\cite{Autoware}, which are often not vulnerabilities.


\subsubsection{ICV security}
\vspace{-1mm}
Beyond security concerns specific to ADSs, several studies have investigated broader attack surfaces and threats in ICVs.
Pekaric et al.~\cite{pekaric2021taxonomy} conducted a literature review and identified 48 types of attacks from existing papers, categorized by access levels—physical, close proximity, and remote—and further subdivided them into targets such as ECUs, radio systems, and Wi-Fi. Similarly, Sommer et al.~\cite{sommer2019survey} proposed a simpler classification based on security threats, including spoofing, tampering, repudiation, information disclosure, denial of service (DoS), and privilege escalation.

Thing et al.~\cite{7917080} presented a taxonomy covering both physical access attacks (e.g., code modification and injection) and remote attacks (e.g., signal spoofing and jamming). Sun et al.~\cite{9447840} categorized cybersecurity risks into in-vehicle network attacks, V2X communication attacks, and other threats, including GPS spoofing, CAN bus vulnerabilities, DoS, and replay attacks. Limbasiya et al.~\cite{LIMBASIYA2022100515} surveyed attack detection and prevention techniques across categories such as impersonation, DoS, Sybil, relay, injection, and side-channel attacks. Gupta et al.\cite{10226207} classified ICV attacks from software, hardware, and network perspectives, identifying common types such as injection, DoS, and side-channel attacks.

Niroumand et al.~\cite{10500196} provided a taxonomy of attacks and mitigation strategies, treating the vehicle as a control system and focusing on attacks like DoS, jamming, man-in-the-middle, and data injection. Bouchelaghem et al.\cite{10.1007/978-3-030-68887-5_15} proposed a taxonomy of attack surfaces in autonomous vehicles and reviewed recent real-world attack experiments, covering six major categories. Luo et al.~\cite{luo2022cybersecurity} conducted a systematic review of automotive cybersecurity testing, with a focus on testing methodologies and tools.

While these works offer valuable insights into ICV threats and taxonomy construction, they primarily rely on literature reviews, expert analysis, or industry interviews. As such, they remain largely conceptual and lack empirical validation through real-world exploits. In this paper, we aim to unify the findings from these studies into a comprehensive taxonomy and validate each category using a dataset of confirmed, real-world exploitable vulnerabilities.

\subsubsection{Bug studies in other domains}
Numerous empirical studies have analyzed bugs across various software systems, focusing on their symptoms, root causes, and broader characteristics~\cite{9283969,9401981,10.1145/3650212.3680366,10.1145/3468264.3468591,10.1145/2931037.2931074,REZAEINASAB2023111563,jing2024revisiting}. These include analyses of bugs in compilers (e.g., GCC~\cite{10.1145/2931037.2931074}), defects in deep learning frameworks~\cite{10.1145/3468264.3468591}, and issues in container runtimes~\cite{10.1145/3650212.3680366}.
Our work shares a similar goal: to systematically characterize and evaluate vulnerabilities, specifically in the ICV domain, based on large-scale real-world exploits. 
}

\begin{figure*}[t]
\centering
\includegraphics[width=1.00\linewidth]{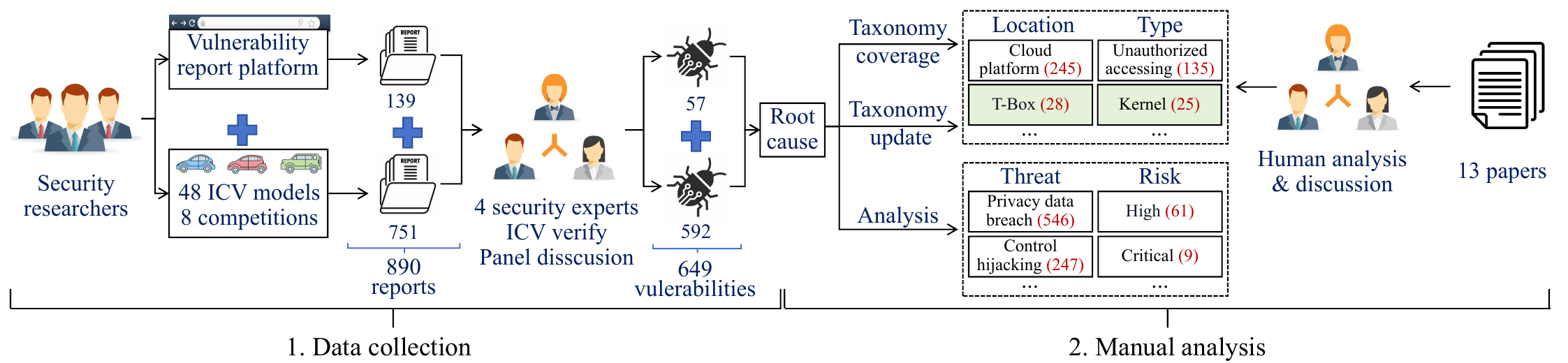}
\vspace{-6mm}
\caption{\revised{The overview of our methodology.}}
\label{fig:methodology}
\end{figure*}

\section{Methodology}\label{sec: data}
To systematically understand the security status of ICVs and identify their primary threats, a comprehensive analysis of ICV vulnerabilities is needed.
This process, as illustrated in Fig.~\ref{fig:methodology}, consists of two main stages: 
data collection and manual analysis.
First, we collected a total of 890 vulnerability reports through an online reporting platform, including 139 reports submitted daily by researchers and 751 reports from vulnerability discovery competitions.
Each report underwent rigorous verification and was tested on real vehicles, ultimately confirming 649 reproducible vulnerabilities.
\revised{Second, we collected 13 representative papers related to ICV security, most of which are survey papers that themselves summarize a broad range of prior ICV research. Since these papers focus on different aspects of ICV security, we manually extracted and analyzed their taxonomies in terms of vulnerability locations and types, and unified them into a consolidated taxonomy.
Third, we mapped the collected vulnerabilities to these existing taxonomies to evaluate their coverage and identify gaps, allowing us to update the taxonomy with newly discovered categories. In addition, we classified the vulnerabilities by threat types and risk levels to provide a more comprehensive understanding of their security implications.}


\subsection{Data collection}\label{sec: vul_collect}
To ensure the comprehensiveness and diversity of the dataset, we collect ICV vulnerabilities through an online vulnerability reporting platform, including vulnerabilities submitted daily by researchers as well as those from ICV vulnerability discovery competitions.

\subsubsection{ICV vulnerability reporting platform}
We established and operated an online platform for vulnerability reporting and tracking.
This platform enables security researchers to submit vulnerabilities and helps manufacturers monitor the security status of their products in real time.
During this study, the platform received 139 vulnerability reports submitted daily by researchers.
After verification, 57 vulnerabilities were confirmed to meet the following criteria: reproducible, within the scope of ICV security, non-duplicative, and validated through testing on real vehicles.

\subsubsection{ICV vulnerability discovery competitions}
To comprehensively collect ICV vulnerabilities, we organized 8 vulnerability discovery competitions between January 2023 and April 2024 in China, collectively referred to as the ``Anonymous Cup''.

\revised{
\textbf{Ethical consideration.}  We carefully addressed the potential ethical concerns related to the competitions. All activities were conducted in accordance with strict ethical standards: the use of ICVs was fully authorized by their respective manufacturers; participants signed non-disclosure agreements; and all confirmed vulnerabilities were responsibly disclosed to the manufacturers for remediation prior to any public release. These measures ensured minimal security risk and full compliance with legal and ethical guidelines.}

Each competition lasted approximately 15 days and involved 6 different vehicle models, with a total of 48 ICVs participating.
\revised{These ICVs represented 19 manufacturers, including brands such as Bavarian Motor Work (BMW), Tesla, and Audi.}
To protect \revised{confidentiality}, specific brands and models were not disclosed.
The ICVs were either directly provided by manufacturers or rented through local rental companies.
The average daily rental cost was \$71, with a total expense of approximately \$51,000.
All ICVs supported network connectivity, and 34 of them (70.8\%) were equipped with a remote control APP. 

\revised{
The competition consisted of two phases: a testing phase and a demonstration phase. During the testing phase, participants adopted a black-box testing approach—no detailed documentation or configuration information was provided. Each team was granted legitimate access to a test vehicle and its associated mobile application, simulating a typical penetration testing environment. Some participants also employed social engineering techniques (e.g., gathering publicly available information) to aid in their reconnaissance.
In the subsequent demonstration phase, participants were required to reproduce their attacks on a separate vehicle of the same make and model without any legitimate access, thereby simulating a real-world attack scenario targeting an unsuspecting victim.
}
%

Participants submitted vulnerability reports via an online platform.
Across the eight competitions, a total of 751 submissions were received.
To minimize data redundancy among the vehicles, deduplication was based on OS versions (identified through the IVI interface) and underlying hardware models (determined via post-competition firmware extraction).
For vulnerabilities that could not be verified initially, submitters were required to provide additional information until confirmation.

\begin{figure}[t]
\centering
\includegraphics[width=0.88\linewidth]{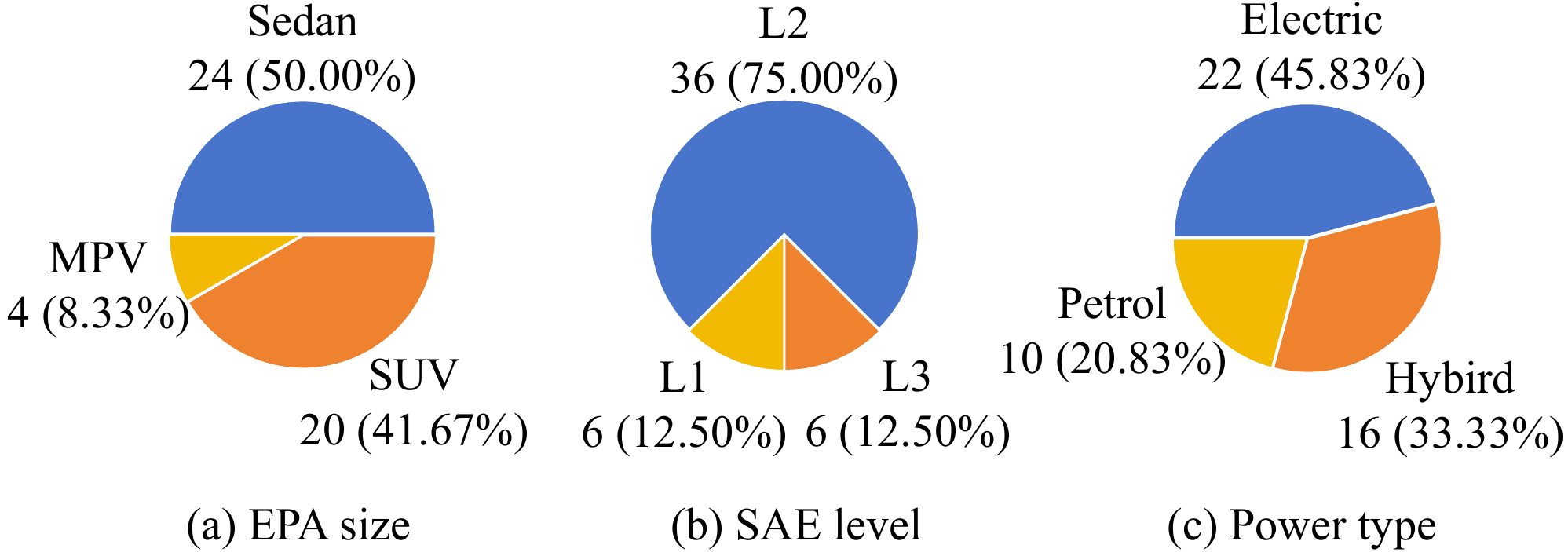}
\vspace{-2mm}
\caption{\revised{The distribution of ICV classes.}}
\label{fig:threepie}
\vspace{-4mm}
\end{figure}

\vspace{-3mm}
\revised{
\subsubsection{Profile of the participating ICVs and participants}\label{sec: icv profile}
Due to confidentiality and business concerns, we are unable to release detailed information about the vulnerabilities and the specific participating ICVs. However, to enhance the reproducibility of our study and provide better context for the test subjects, we present the profile of the participating ICVs.
Specifically, we classify the participating vehicles along four dimensions:
First, based on the widely adopted classification standard from the U.S. Environmental Protection Agency (US EPA~\cite{us_epa_car_classification}), the vehicles are categorized into sedans (50.00\%), standard SUVs (41.67\%), and MPVs (8.33\%), as shown in Fig.~\ref{fig:threepie} (a).
Second, according to the SAE levels of driving automation~\cite{sae_j3016_update}, the vehicles span Levels L1, L2, and L3, with Level 2 being the most common, accounting for 75\% of the total, as illustrated in Fig.~\ref{fig:threepie} (b).
Third, by power type, the vehicles include electric vehicles (45.83\%), petrol vehicles (20.83\%) and hybrids (33.33\%), as shown in Fig.~\ref{fig:threepie} (c).
Finally, in terms of country of manufacture, the vehicles come from six countries, with the top three being China (58.33\%), the United States (12.5\%), and Germany (12.5\%).

Most participants were from China, as the competitions were held locally.
They primarily came from cybersecurity companies and academic institutions, bringing strong technical foundations and extensive hands-on experience.
Based on registration data, approximately 95\% had prior experience in national or enterprise-level security competitions, and over 85\% had more than three years of experience in ICV security research.
Although we did not conduct a formal skill survey prior to the competitions, we performed a correlation analysis between participants’ technical backgrounds and the vulnerabilities they submitted.
This helps support reproducibility in the context of human-factor studies.

To facilitate further analysis, we have released all raw data in our dataset~\cite{dataset}. This includes the ICV ID, participant ID, primary skills used for each vulnerability, and high-level characteristics of each ICV (e.g., EPA category, SAE level, power type). These relationships enable deeper analysis, such as exploring how participant skill sets relate to vulnerability types or how vulnerabilities correlate with vehicle characteristics (e.g., EPA classification). More detailed discussions can be found in Section~\ref{vehicletaxonomy}.}

\subsubsection{Existing ICV vulnerabilities.}
\vspace{-2mm}
To better understand existing ICV vulnerabilities, we collected 13 recent papers~\cite{pekaric2021taxonomy, pham2021survey, sommer2019survey, 7917080, 9447840, LIMBASIYA2022100515, 10226207, 10500196, 10.1007/978-3-030-68887-5_15, luo2022cybersecurity, 9625017, tang2023survey, 9284001} that focus on ICV attacks and taxonomy construction. Most of these are survey papers that summarize a wide range of relevant literature in the field. We manually extracted the reported attacks and vulnerability types from each paper to form a consolidated view.
The detailed list of vulnerability types extracted from each source is available on our project website~\cite{detype}.

\subsection{Manual analysis}

\revised{Based on the vulnerabilities and attacks extracted from each paper, we manually unified them into two taxonomies: vulnerability locations and vulnerability types.
This task was relatively straightforward, as many of the attacks were conceptually similar across papers.
In general, the taxonomy draws upon established standards like CWE~\cite{cwe}, while adapting them to the specific context of ICVs.
In cases where different terminologies were used to describe similar attacks, we grouped them under the same unified category.
The taxonomy construction process was conducted collaboratively by three security experts.
Any disagreements were resolved through discussion until a consensus was reached.
The details of the unified taxonomies are presented in Section~\ref{sec:vul_loc_tax} and Section~\ref{sec:vul_type_tax}.}

\vspace{-6mm}
\revised{
\subsection{Mapping of existing vulnerabilities}
We further analyzed the root causes of the collected vulnerabilities and mapped them to the existing taxonomies.
}
The analysis was conducted by four security experts, each with over five years of experience in ICV vulnerability auditing and competition judging.
The analysis followed a collaborative, discussion-based approach to ensure rigor and reliability.
The analysis was based on 890 vulnerability reports from 48 ICVs, among which 649 were confirmed as valid vulnerabilities. 
Each vulnerability was independently analyzed by the experts to determine its root cause and draft a description, followed by classification.
In cases of disagreement, the team engaged in multiple rounds of discussion until consensus was reached, ensuring consistency and accuracy in classification results.

Our classification followed a multi-dimensional framework: \ding{182} \revised{\textbf{Location-based categorization}: To analyze the distribution of vulnerabilities, we first mapped them to the unified location-based taxonomy.} \ding{183} \revised{\textbf{Type-based categorization}: To support targeted mitigation strategies, we mapped each vulnerability to the unified type-based taxonomy.} \ding{184} \textbf{Threat-based categorization}: To identify potential security consequences, we further classified vulnerabilities into six threat categories, including privacy data breach and control hijacking. \ding{185} \textbf{Risk-based categorization}: To guide remediation priorities, we assigned each vulnerability a risk level (critical, high, medium, or low) based on its potential impact.

\revised{For vulnerabilities that could not be mapped to any existing categories, we proposed new categories and incorporated them into the updated taxonomy.}
This data-driven, expert-led, multi-dimensional classification framework provides a systematic overview of real-world ICV vulnerabilities and offers practical guidance for future security research and mitigation efforts.
A total of \revised{98} person-months were invested in organizing the competitions, reviewing submissions, and conducting the vulnerability analysis, involving four security experts and two supporting staff members.

\vspace{-3mm}
\section{Location taxonomy}
\label{sec:vul_loc_tax}

\begin{figure}[t]
\centering
\includegraphics[width=0.94\linewidth]{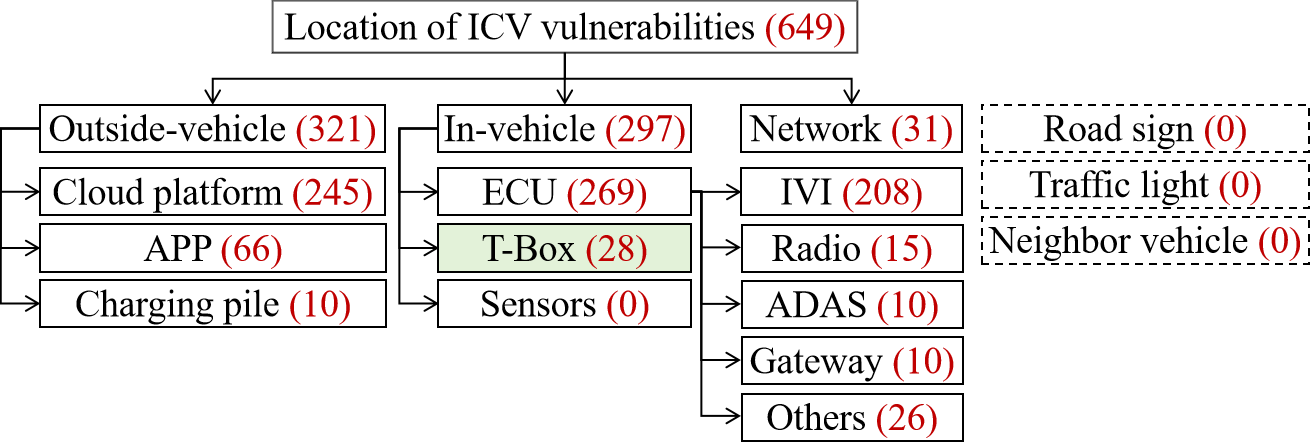}
\vspace{-2mm}
\caption{\revised{Location taxonomy of ICV vulnerabilities.}}
\label{fig:location}
\vspace{-4mm}
\end{figure}

\revised{Fig.\ref{fig:location} presents the location-based taxonomy, which includes 11 leaf categories. Categories in white represent those also covered in previous works, while categories in green are newly discovered in our study based on real-world vulnerabilities and were not previously included. 
Categories with dashed borders fall outside the scope of our analysis since we only consider vulnerabilities in software. 
Each number indicates how many of the collected vulnerabilities fall under that category. We next introduce each category briefly.
}


\subsection{Outside-vehicle}
The outside-vehicle category, which includes external yet security-critical systems such as cloud platforms, APPs, and charging piles, accounts for 321 vulnerabilities and represents 49.5\% of all vulnerabilities in the dataset.

\insightbox{
\revised{\textbf{Finding 1:}
Overall, our collected vulnerabilities demonstrate high coverage, encompassing 90.91\% (10 out of 11) of the considered location categories.
The only category not covered is Sensors.
Additionally, our analysis identified 1 new location category that was not present in prior taxonomies.
We also compared our coverage against the 13 existing papers and found that their individual coverage ranged from 27.27\% to 63.64\%, which is significantly lower than the coverage achieved by our dataset.}
}

\subsubsection{Cloud platform}
The cloud platform has a total of 245 vulnerabilities, accounting for the vast majority (76.3\%) of outside-vehicle vulnerabilities.
As a core support system, the cloud platform is responsible for key functions such as remote control, OTA (On-The-Air) updates, and authentication, involving multiple subsystems and communication protocols, which significantly broadens the attack surface.
Studies show that each ICV connects to an average of 5.1 cloud platforms.
Inconsistent standards among manufacturers, along with weak security design and permission management during integration, further exacerbate the number of vulnerabilities.


\subsubsection{APP}
As the primary outside-vehicle component for remote control, the APP also has certain vulnerabilities, accounting for 20.6\% of all outside-vehicle vulnerabilities.
Our study found that ICV APPs often have serious flaws in data transmission, such as binding control commands only to the Vehicle Identification Number (VIN) without additional encryption, making them susceptible to privacy leaks and replay attacks.
\revised{Note that additional details about the APP-related taxonomy are available. Due to space limitations, the detailed taxonomy information is provided in our dataset~\cite{dataset}.}

\subsubsection{Charging pile}
The charging pile represents a smaller yet notable risk, comprising 3.1\% of outside-vehicle vulnerabilities.
Charging pile systems interact with both the cloud platform and the vehicle to manage authentication, billing, and charging processes, relying on complex protocols such as ISO 15118 and OCPP~\cite{deb2022smart}. 
The openness of these protocols and their connectivity to multiple systems expand the attack surface~\cite{metere2021securing}, where insufficient authentication and encryption can lead to security risks, including information leakage and privilege escalation.

\subsection{In-vehicle}
In-vehicle vulnerabilities total 297, accounting for 45.8\% of the total.
These vulnerabilities are divided into six subcategories: IVI, radio, ADAS, gateway, others, and T-Box, with details as follows.

\subsubsection{IVI}
IVI vulnerabilities account for 70.0\% of in-vehicle vulnerabilities.
To support features such as Bluetooth, Wi-Fi, and voice assistants, the IVI system opens a large number of network ports and debugging interfaces, many of which are not essential for normal operation but provide entry points for attackers.
In addition, the debugging mode may be leaked, and some manufacturers keep debugging ports open for maintenance convenience, increasing security risks.
Moreover, to reduce costs, some manufacturers still use outdated software components and communication protocols, further expanding the attack surface.


\subsubsection{Radio}
Radio vulnerabilities account for 5.1\% of in-vehicle vulnerabilities.
It uses wireless technologies such as RFID or NFC for signal transmission.
Despite encryption, the inherent openness of wireless communication makes it susceptible to interception and replay attacks~\cite{10.1145/3627827}.
Additionally, the proximity-based, contactless keyless entry feature expands the attack surface, particularly making it vulnerable to relay and replay attacks, as it automatically detects keys within a specified range~\cite{ASHWORTH2023100587}.

\subsubsection{ADAS, gateway and others}
This category includes ADAS, gateway, and simpler ECUs such as Bluetooth key signal receivers, with a total of 46 vulnerabilities identified, accounting for 15.5\% of all in-vehicle system vulnerabilities.
Vulnerabilities in ADAS arise from reliance on external sensors, high-precision maps, and real-time data from the cloud platform, with open data interfaces and multi-source data fusion increasing the risk of interference or tampering~\cite{10.1016/j.inffus.2024.102822}. 
Gateway ECU vulnerabilities stem from its role as a central communication hub with open interfaces like firewalls and intrusion detection, which increase the attack surface, and complex security configurations that can lead to misconfigurations or insufficient encryption~\cite{semiconductor2018automotive}. 
Simpler ECUs like Bluetooth key signal receivers vary across different ICV models, lacking common characteristics and presenting diverse security challenges~\cite{salfer2015attack}.

\subsubsection{T-Box}
T-Box vulnerabilities account for 9.4\% of in-vehicle vulnerabilities.
As the core communication interface, it is responsible for receiving and verifying remote control commands.
Its use of multiple communication protocols, particularly MQTT (Message Queuing Telemetry Transport), along with its high-level access privileges, expands the attack surface~\cite{zhang2024endogenous}.
Misconfigurations or leaks in certificate management further exacerbate security risks.

\vspace{2mm}
\insightbox{
\textbf{Finding 2:}
ICV vulnerabilities are pervasive and are strongly influenced by the functions of individual modules.
Outside-vehicle, especially the cloud platform, account for a significant portion of vulnerabilities due to their extensive connectivity and reliance on data.
In-vehicle vulnerabilities are most prominent in the IVI system, driven by its exposure to external data inputs and complex software architecture. 
Although network vulnerabilities are relatively few, they still pose significant risks due to their critical role in vehicle system communication.
}

\subsection{Network}
There are 31 network vulnerabilities, accounting for approximately 4.8\% of the total.
As the core communication backbone of ICV systems, the network component integrates various protocols (such as CAN, Telnet, and SSH) to enable interoperability and communication between functional modules.
However, these protocols lack a unified security design, making transmitted data susceptible to interception and tampering.
In addition, in-vehicle networks (such as the CAN bus) typically trust all connected devices by default, allowing attackers to easily access the system through physical interfaces (such as OBD-II) and launch malicious operations, thereby exposing serious weaknesses in access control and data integrity.

\section{Type taxonomy}
\label{sec:vul_type_tax}

\begin{figure*}[t]
\centering
\includegraphics[width=0.94\linewidth]{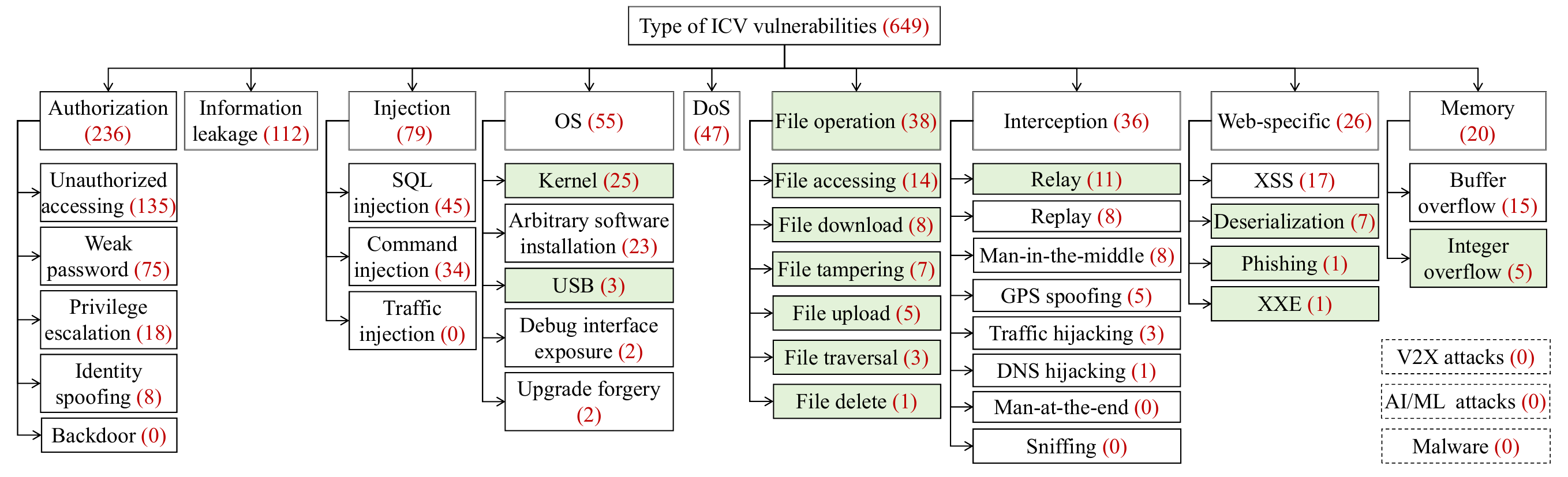}
\vspace{-4mm}
\caption{\revised{Type taxonomy of ICV.}}
\label{fig:type0714}
\vspace{-2mm}
\end{figure*}

\begin{table*}[t]
\centering
\begin{center}
\footnotesize
\setlength{\tabcolsep}{2.1pt}
\caption{\revised{Frequency of vulnerability types for each location.}}
\vspace{-2mm}
\begin{tabular}{c|c|ccc|ccc|ccc|ccc|ccc|ccc|ccc|ccc|c|c}
\toprule
\label{tab:matrix}
\multirow{2}{*}{Type}  & \multirow{2}{*}{Sub-Type} & \multicolumn{3}{c|}{Cloud platform} & \multicolumn{3}{c|}{IVI} & \multicolumn{3}{c|}{APP} & \multicolumn{3}{c|}{ECU} & \multicolumn{3}{c|}{Network} & \multicolumn{3}{c|}{T-Box} & \multicolumn{3}{c|}{Radio} & \multicolumn{3}{c|}{Charging pile} & \multicolumn{2}{c}{\multirow{2}{*}{Sum}}\\
\cmidrule{3-26}
& & \#N & T & R & \#N & T & R & \#N & T & R & \#N & T & R & \#N & T & R & \#N & T & R & \#N & T & R & \#N & T & R &\multicolumn{2}{c}{}\\
\midrule
\multirow{4}{*}{\ding{182} Authorization} &Unauthorized accessing & 57 & \ding{172}\ding{177} & C & 45 & \ding{172}\ding{173} & H & 19 & \ding{172}\ding{173} & C &1 & \ding{172} & L &6 &  \ding{172}\ding{173} & H &7 & \ding{173} &C &-&- & -  & -& -& -&135& \multirow{4}{*}{236}\\
 &Weak password & 31 & \ding{172}\ding{177} & C &14 & \ding{172}\ding{173} & H &2 & \ding{172}\ding{173} & L &7 & \ding{172}\ding{173} & M &15 & \ding{172}\ding{173} & L&6 & \ding{172}\ding{173} &M &-&- &- & -& -& -&75 \\
 &Privilege escalation & - & - &- &16 &\ding{172}\ding{173} & H &- & - &- &1 &\ding{173} & M &- &-&- &1 & \ding{173}& M & -&- & -& -& -& -&18\\
 &Identity spoofing & 7 & \ding{172}\ding{177} & C &- & - &- & 1 & \ding{172}\ding{173} & L &- & -&- &- & - &- &- & -&- &- & - &-& -& -& -&8 \\
\midrule
\multirow{1}{*}{\ding{183} Information leakage}&- & 53 & \ding{172}\ding{177} & H &13 & \ding{172}& H &27& \ding{172}& H &7& \ding{172} & L &1& \ding{172} & L &3 & \ding{172} & M &-&- & - &8 &\ding{172} &L & 112 & 112\\
\midrule
\multirow{2}{*}{\ding{184} Injection} &SQL injection & 31 & \ding{172} & H &11 & \ding{172} & M &1 & \ding{172} & L &- & - &-  & - &- & - &-& - &- & - &- & - & 2& \ding{172}& M &45&\multirow{2}{*}{79}\\
 &Command injection & 16 & \ding{172}\ding{177} & H &5 & \ding{172}\ding{173} &H & 6 & \ding{172}\ding{173} & M &3 & \ding{172} & M &- & -  & -&4 & \ding{173} & H&- & -& -& -& -& -&34\\
\midrule
\multirow{5}{*}{\ding{185} OS}&Kernel & - & -& -&25 &\ding{172}\ding{173} & M &- &- &- & - &- & - &- & - &- & - & - &- & - & - &- & -& -& -&25& \multirow{5}{*}{55}\\
 &Arbitrary software installation & 1 & \ding{172}\ding{177} & M &17 & \ding{172}\ding{173} & H &- & -& - &5 & \ding{176} & M &- & -& -&- & - & -&- & -&-& -& -& -&23\\
 &USB & - & - & -&3 & \ding{172}\ding{173}& L&- & -& -&- & -& -&- & -& -&- & -& -&- & -& -& -& -& -&3\\
 &Debug interface exposure & - & -& - &1 & \ding{172}\ding{173}& M &- & -&- & -&- & -&1 & \ding{173}& M&- & -&- & -&- & -& -& -& -&2\\
 &Upgrade forgery & - & -  &- & -&- & -&- & - & -&1 & \ding{173} & H&- & -& -&1 &\ding{173}& M &-&-&-& -& -& -&2\\
\midrule
\multirow{1}{*}{\ding{186} DoS} & - & 1 & \ding{174} & L &26 & \ding{174} & L &7  & \ding{174} & L &5 & \ding{174} & H &5 & \ding{174} & M &1 & \ding{174} & M&2 & \ding{174} & M & -& -& -& 47 & 47\\
\midrule
\multirow{6}{*}{\ding{187} File operation} &File accessing & 7 & \ding{172}& H &7 & \ding{172}& L &- & - &- & -&- & - &- &- &- &- &- &- &- & -& -& -& -& -&14& \multirow{6}{*}{38} \\
 &File download & 7 &\ding{172}& M &- & -&- &- &- &-&1& \ding{172} & L &- & - &-&- &- &- &- & -&- & -& -& -&8\\
 &File tampering & - & -  & -&7 & \ding{172}\ding{173}& M&- & - & -&- & - & -&- & -& - &- & -& - &- & -& - & -& -& -&7 \\
 &File upload & 4 &\ding{172}\ding{177} & M &- & -&- &- & -&- &- & -&- &1 & \ding{173} & M &- & - &- &- &-&-& -& -& -&5\\
 &File traversal & 3 & \ding{172} & L&- & -& - &- & -& -&- & -& -&- & -& -&- & -& -&- & -& -& -& -& -&3\\
 &File delete & 1 & \ding{174}& L& - &- & -&- &- & -&-&- & -&- &- &- &- &- &- &- &- &- & -& -& -&1\\
\midrule
\multirow{6}{*}{\ding{188} Interception}&Relay & - & - & -&- & - & -&- & -& - &2 &\ding{175}& M &1&\ding{173} & M&- & -&-&8 & \ding{175}& H & -& -& -&11 & \multirow{6}{*}{36}\\
 &Replay & - & -& -  &- & -& - & 2 &\ding{172}\ding{173} & H &- & - &- & 1 & \ding{174} & L &- & - &- &5 & \ding{175} & C & -& -& -&8\\
 &Man-in-the-middle & - & - & - &7 &\ding{172}\ding{173}& M &1& \ding{172}\ding{173} & M &- & -& -  &- & - & - &- & -& -&- & - & - & -& -& -&8 \\
 &GPS spoofing & - & -  & -& - &- & -&- & -  & -&5 &\ding{174}\ding{176}&L & - & -&- & -  & -&- &- & - & -  & -& -& -&5\\
 &Traffic hijacking & - & -  & -&3  & \ding{172}\ding{173} &M &- & -& - &- & -& - &- & - & -&- & -& -&- & - & -& -& -& -&3\\
 &DNS hijacking & - & -  & -&1&\ding{172}\ding{173} & M &- & -&- & -&- &- &- & -&- &- & -&-  &- & -&-& -& -& -&1\\
\midrule
\multirow{4}{*}{\ding{189} Web-specific} &XSS & 17 & \ding{172} & L&-&- &- &- &- &- &- & - &- &- & - &-&- & - &-&- &-&-  & -& -& -&17&\multirow{4}{*}{26}\\
 &Deserialization & 7 & \ding{172}\ding{177} & M &- &- &-&- & -  &- & - &- & - &- & - &- & -  & - & - & -  & - & - & -& -& -&7 \\
 &Phishing & 1 & \ding{172}& L&- & -&- &- &- &- &- &- &- &- &- & -&- &- &- &- &- &- & -& -& -&1\\
 &XXE & 1 & \ding{172} & L&- & -&- &- &- &- &- &- &- & -&- &- &- & - &- &- &- &-& -& -& -&1\\
\midrule
\multirow{2}{*}{\ding{190} Memory} &Buffer overflow & - & - &-&4 & \ding{172}\ding{173} & M &- &-&-  &6 & \ding{173} & H &- &-&- &5 &\ding{173} & L &- & -&- & -& -& -&15 & \multirow{2}{*}{20}\\
 &Integer overflow & - & - & -&3 &  \ding{172}\ding{173}& H &- & - & -&2 & \ding{172}\ding{173}&H &- & -& - &- & - & -&- & -& - & -& -& -&5\\
\midrule
\multicolumn{2}{c|}{Sum} & \multicolumn{3}{c|}{245} & \multicolumn{3}{c|}{208} & \multicolumn{3}{c|}{66} & \multicolumn{3}{c|}{46} & \multicolumn{3}{c|}{31} & \multicolumn{3}{c|}{28}& \multicolumn{3}{c|}{15} & \multicolumn{3}{c|}{10}& \multicolumn{2}{c}{649} \\
\bottomrule
\end{tabular}
\end{center}

\begin{flushleft}
\footnotesize
Threat (T):
\ding{172} Privacy data breach;
\ding{173} Control hijacking;
\ding{174} System damage;
\ding{175} Unauthorized unlocking;
\ding{176} Driving accident;
\ding{177} Ransomware;\\
Risk (R):
Low (L);
Medium (M);
High (H);
Critical (C);\\
Note:
The ``Risk'' column indicates the maximum possible risk level of vulnerabilities for this type in this location.
\end{flushleft}

\vspace{-2mm}
\end{table*}
To facilitate the development of targeted detection techniques to address the diverse security challenges in ICVs, we characterize the vulnerabilities based on their types.
\revised{
As shown in Fig.~\ref{fig:type0714}, we identified a total of 35 categories in the vulnerability type taxonomy. Similar to the location-based taxonomy, 22 of these categories (shown in white) are already covered by existing works. 
In addition, we discovered 13 new categories (in green) that were not previously reported. 
Three categories were marked as out-of-scope, either because they relate to autonomous vehicle components or involve malware types that typically require user interaction.
Below, we will introduce these types.
}


\textbf{\ding{182} Authorization vulnerability.} Table~\ref{tab:matrix} shows that authorization vulnerabilities are the most common type, with a total of 236 cases, accounting for 36.4\% of all vulnerabilities. 
These vulnerabilities are often caused by misconfigurations or design flaws and can lead to high-risk issues such as privacy breaches.
Specifically, they include unauthorized accessing (135 cases), weak password (75 cases), privilege escalation (18 cases), and identity spoofing (8 cases).
These vulnerabilities are mainly found in external modules (e.g., cloud platform) and in-vehicle modules (e.g., IVI), indicating significant security risks in these components due to inadequate authorization management.

\textbf{\ding{183} Information leakage vulnerability.} There are 112 information leakage vulnerabilities, accounting for 17.3\% of the total.
They are mostly caused by implementation errors, which can easily lead to the leakage of sensitive information such as the VIN and keys.
Since external modules like cloud platforms and APPs frequently handle large amounts of sensitive data, information leakage vulnerabilities are particularly common in these modules.

\textbf{\ding{184} Injection vulnerability.}
There are 79 injection vulnerabilities in total, accounting for 12.2\% of all cases, mainly including SQL injection (45 cases) and command injection (34 cases).
SQL injection typically results from insufficient input validation, allowing malicious code to be executed and causing privacy data breaches.
\revised{We categorize it as a general injection vulnerability rather than limiting it to web scenarios, as it also appears in non-web contexts such as IVI systems and APPs.}
Command injection occurs due to inadequate validation of external inputs, enabling attackers to execute unauthorized commands.
As shown in Table~\ref{tab:matrix}, injection vulnerabilities are mainly found in cloud platforms and IVI systems, which frequently handle user inputs and database operations, making them more susceptible to attacks.

\textbf{\ding{185} OS vulnerability.}
OS vulnerabilities are caused by security flaws at the system level, with a total of 55 cases, accounting for 8.5\% of all vulnerabilities.
They are mainly categorized into five types: kernel vulnerabilities (25 cases), arbitrary software installation (23 cases), USB vulnerabilities (3 cases), debug interface exposure (2 cases), and upgrade forgery (2 cases).
\revised{Among them, kernel vulnerabilities and USB vulnerabilities are unique to our vulnerability type taxonomy.}
As the most common type of OS vulnerability, kernel issues can lead to privacy breaches or control hijacking, with risk levels reaching up to medium.
These vulnerabilities are primarily found in in-vehicle modules, especially IVI systems, which are particularly sensitive to kernel vulnerabilities and malicious software installations.

\textbf{\ding{186} DoS vulnerability.}
There are 47 DoS vulnerabilities, accounting for 7.2\% of the total vulnerabilities (see Table~\ref{tab:matrix}).
These vulnerabilities are primarily found in IVI components, making up 55.3\% of all DoS vulnerabilities.
The root causes include component configuration errors, kernel defects, and logical design weaknesses, such as protocol stack vulnerabilities or resource exhaustion attacks, which may lead to system damage within the affected components.

\textbf{\ding{187} File operation vulnerability.}
File operation vulnerability is a new high-level category in our classification, with a total of 38 cases, accounting for 5.9\% of all vulnerabilities.
This category includes subtypes such as file accessing (14 cases), file download (8), file tampering (7), file upload (5), file traversal (3), and file deletion (1).
As shown in Table~\ref{tab:matrix}, these vulnerabilities are primarily found in cloud platforms and IVI systems.
Cloud platforms, which handle large amounts of file data, are vulnerable to insufficient access control, while IVI systems, which frequently perform local file operations, are exposed to risks such as tampering and unauthorized downloads.

Among them, file accessing vulnerabilities are typically caused by improper partition management in cloud platforms or incorrect permission configurations in IVI systems, allowing attackers to read sensitive files and cause privacy breaches~\cite{10.1145/3343737.3343753}.
\revised{File download vulnerabilities allow attackers to obtain sensitive files by modifying URLs or exploiting unvalidated download paths, while file traversal vulnerabilities bypass directory restrictions by crafting malicious paths to access unintended files.
Although all three involve unauthorized access, they differ in technical mechanisms, attack objectives, and stages in the attack chain, and thus represent three distinct types of vulnerabilities.}

\textbf{\ding{188} Interception vulnerability.}
Table~\ref{tab:matrix}, there are 36 interception vulnerabilities in total, accounting for 5.5\% of all vulnerabilities.
It include six subtypes: relay (11 cases), replay (8), man-in-the-middle (8), GPS spoofing (5), traffic hijacking (3), and DNS hijacking (1).
Interception vulnerabilities are primarily concentrated in IVI, ECU, and radio modules.
IVI and ECU, which are involved in internal vehicle communication, are vulnerable to interception attacks, while network modules like T-Box and radio, due to their wireless communication characteristics, are also susceptible to signal hijacking.
\revised{Among these vulnerabilities, relay attacks are a vulnerability type unique to our classification.}
These vulnerabilities typically arise from the system's failure to encrypt or authenticate transmitted information, allowing attackers to exploit relay communication to perform unauthorized actions, potentially even unlocking the vehicle without authorization~\cite{francillon2011relay}, with risk levels reaching High.

\textbf{\ding{189} Web-specific vulnerability.}
As shown in Table~\ref{tab:matrix}, there are 26 web-specific vulnerabilities, accounting for 4.0\% of the total.
These vulnerabilities are categorized into four types: XSS (17 cases), deserialization (7), phishing (1), and XXE (1).
Web-specific vulnerabilities are only found in the cloud platform, indicating that such issues are mainly confined to the cloud environment.
XSS vulnerabilities arise from the failure of the manufacturer's cloud platform to properly handle user inputs (e.g., lack of input validation), allowing attackers to inject malicious scripts that are executed through other users’ browsers, leading to privacy data breaches.
\revised{The remaining vulnerability types, such as deserialization, phishing, and XXE, are unique to our classification.}

\textbf{\ding{190} Memory vulnerability}
A total of 20 memory vulnerabilities were identified, which can be further divided into two types: buffer overflow (15 cases) and integer overflow (5 cases).
Buffer overflow vulnerabilities are mainly caused by developers failing to properly handle user inputs or data, allowing attackers to trigger data overflows and corrupt memory.
Integer overflow, \revised{a vulnerability type unique to our classification,} arises from inadequate consideration of data type boundaries, which may lead to memory corruption and subsequently cause privacy breaches, with risk levels reaching high.
Memory vulnerabilities primarily occur in IVI and ECU modules, as these modules involve complex data processing and are prone to memory errors due to insufficient input validation or improper resource management.

\insightbox{
\revised{\textbf{Finding 3:}
The identified vulnerabilities demonstrate high coverage, spanning 88.57\% (31 out of 35) of the considered type categories. The categories not covered include Backdoor, Traffic injection, Man-at-the-end, and Sniffing. Compared to prior works, our analysis uncovered 13 new vulnerability types. 
We also compared our taxonomy coverage with the 13 existing papers and observed that their individual type coverage ranged from 11.43\% to 40.00\%, further underscoring the comprehensiveness of our dataset.
}
}

\vspace{2mm}
\insightbox{
\revised{\textbf{Finding 4:}
The newly identified vulnerability locations and types highlight the diversity and complexity of the real-world ICV vulnerabilities we collected, underscoring the importance of data-driven analysis as a complement to existing conceptual, literature-based approaches.
Such empirical analysis enables the refinement and expansion of current vulnerability  taxonomies.
}}

\vspace{2mm}
\insightbox{
\textbf{Finding 5:}
Authorization vulnerabilities (36.4\%) and information leakage vulnerabilities (17.3\%) are the most common, reflecting systemic flaws in access control and data protection within ICV systems.
In addition, different modules face distinct types of vulnerabilities (all 26 web-specific vulnerabilities are found in the cloud platform), highlighting the urgency of implementing targeted vulnerability detection strategies.
}
\vspace{2mm}

\section{Threat and risk taxonomy}

To identify specific security consequences and support threat modeling, we categorize vulnerabilities by threat, producing 6 threat categories, as shown in Table~\ref{tab:matrix}.
Among them, privacy data breaches refer to unauthorized access to sensitive personal or vehicle data, compromising confidentiality;
control hijacking occurs when attackers gain unauthorized control over critical vehicle functions, such as steering or braking, endangering driving safety;
system damage refers to malfunctions or disruptions causing temporary or permanent loss of vehicle functionality, impacting operational safety;
unauthorized unlocking enables attackers to access or unlock vehicles without permission, increasing theft risks;
driving accidents result from system manipulation leading to crashes or unsafe driving conditions, posing risks to occupants and others;
ransomware involves malicious software locking vehicle systems or data, demanding payment for restoration and disrupting functionality.
These diverse categories underscore the severe security threats in ICVs, necessitating targeted strategies.

\begin{figure}[t]
\centering
\includegraphics[width=\linewidth]{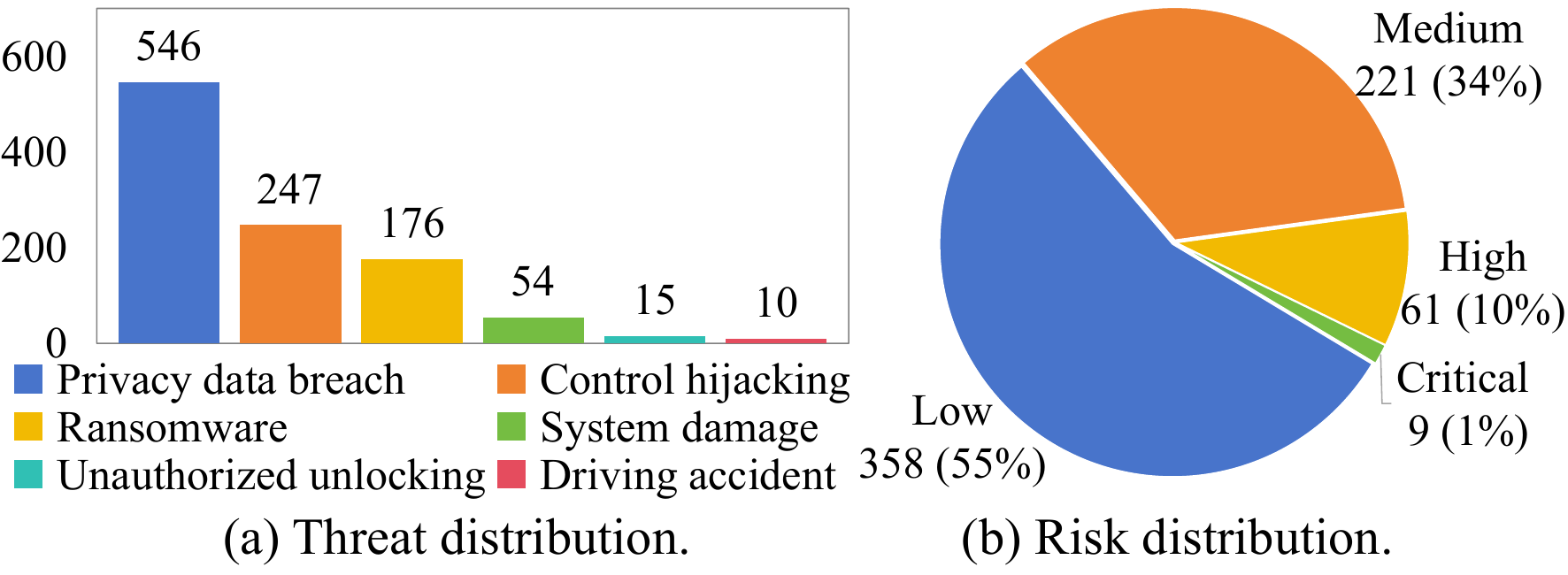}
\vspace{-4mm}
\caption{The threat and risk distribution.}
\label{fig:threat}
\end{figure}

Some vulnerabilities may lead to multiple consequences. For example, certain control hijacking vulnerabilities may also result in privacy breaches.
As shown in Fig.~\ref{fig:threat}-(a), 84.13\% of vulnerabilities involve privacy data leakage, 38.06\% can lead to control hijacking, and 27.11\%, 8.32\%, 2.31\%, and 1.54\% are related to ransomware , system damage, unauthorized unlocking, and driving accidents, respectively.
These multifaceted risks highlight the severe security challenges faced by ICVs and underscore the urgent need for targeted defense strategies.

To prioritize remediation based on impact severity, we classify vulnerabilities by risk, resulting in 4 risk levels, such as low (L), medium (M), high (H), and critical (C).
Specifically, we measures the severity of vulnerabilities using two dimensions: the ease of conducting an attack and the impact of the attack.
A relative scoring system from 1 to 4 is applied, and the risk level is determined based on the vulnerability risk level determination table.
The scoring of each vulnerability is completed through collaboration between experts and the submitter, consisting of three steps: initial scoring by the submitter, expert review, and final consensus, ensuring fairness and accuracy.
For example, if both the ease of conducting the attack and the impact score 4, the vulnerability is classified as critical.
As shown in Fig.~\ref{fig:threat}-(b), 55.16\%, 34.05\%, and 9.34\% of the vulnerabilities were rated as low, medium, and high risk respectively.
The remaining 9 vulnerabilities were rated as critical risk and distributed across cloud platform (4), radio (3), IVI (1), and T-Box (1). 
They are classified as critical mainly because their impact extends to all models of a specific brand or model or affects more than 300,000 vehicles.

\vspace{-8mm}
\revised{
\section{Participants' skills and ICV types}\label{vehicletaxonomy}

To further support potential reproduction, primarily dependent on both the participants who discovered the vulnerabilities and the ICV types in which the vulnerabilities were found, we provide additional analysis of the participant profiles and vehicle characteristics.

\vspace{-2mm}
\subsection{Participants' skills}

As discussed in Section~\ref{sec: icv profile}, although we did not explicitly collect participants' skills, we provided their expertise by analyzing the relationships between participants, the vulnerabilities they discovered, and the technical skills required to uncover those vulnerabilities. Table~\ref{tab: skillplus} presents the overall mapping between major vulnerability locations/types and the corresponding skills needed.

\begin{table}[t]
\centering
\footnotesize
\setlength{\tabcolsep}{3.28pt}
\renewcommand{\arraystretch}{1}
\caption{\revised{Mapping results of the skill set.}}
\begin{tabular}{c|c|cccccccccc}
\toprule
\multicolumn{2}{c|}{Location \& Type} & \ding{172} &\ding{173}&\ding{174}&\ding{175}&\ding{176}&\ding{177}&\ding{178}& \ding{179} &\ding{180}&\ding{181} \\
\midrule
\multirow{8}{*}{Location}& Cloud platform  & - & - &243& -& - &32&-&2&-&-\\	
&IVI &146&146&-&135 & 1 &13&11&9&39&5\\
&APP &63&63&-&-&59&-&63&-&-&-\\
&ECU &19&19&-&19&5&12&-&13&6&3\\
&Network&7&7&-& 7&1&15&-& 15& -&7\\
&T-Box &21&21&-&21&-&6&-&7&-&-\\
&Radio &-&-&-&-&14&1&-&1&-&-\\
&Charging pile&4&4&6&-&-&-&4&-&-&-\\ 
\midrule
\multirow{9}{*}{Type}&Authorization& 90& 90 & 94& 68 & 23 & 72 &  22 & 37 & 1 & 4 \\
&Information leakage & 49& 49 & 58 & 21 & 28 & 1 & 28 & 3 & -  & 1  \\
&Injection & 32& 32 & 47 & 13 & 7 &-&19&-&-&-\\
&OS & 23 & 23 & 1 & 22 & - &3 & 1 & 4 & 27 & - \\
&DoS & 27 & 27 & 1 & 20 & 1 &1 & 7 & 1 & 9  & 8 \\
&File operation & 13 & 13& 22 & 13 & - &- &- &- & 2&1\\
&Interception & 13& 13 & - & 12 & 21 &- &1 &- &1 &1\\
&Web-specific & - & - & 26 & - &-&-&-&-&-&-\\
&Memory & 13 & 13 & - & 13 & - & 2&-&2&5&-\\
\midrule
\multicolumn{2}{c|}{Sum}&260&260&249&182& 80 &79&78&47&45&15 \\

\bottomrule
\end{tabular}

\begin{flushleft}
\footnotesize
\ding{172} Reverse engineering;
\ding{173} Code audit;
\ding{174} Penetration testing; 
\ding{175} Firmware extraction;
\ding{176} Traffic analysis;
\ding{177} Brute force; 
\ding{178} Unpacking; 
\ding{179} Port scanning; 
\newline
\ding{180} Vulnerability reproduction;
\ding{181} Fuzzing;
\end{flushleft}

\label{tab: skillplus}

\end{table}


Overall, we identify 10 categories of technical skills involved in the discovered vulnerabilities. Among them, reverse engineering, code audit, and penetration testing are the most frequently used, with 260, 260, and 249 instances, respectively.

For outside-vehicle components, the dominant skill required for vulnerabilities in the cloud platform is penetration testing (243 occurrences), reflecting the need for in-depth analysis of communication protocols and business logic. Vulnerabilities in APPs typically involve a combination of reverse engineering (63), code audit (63), and unpacking (63), with occasional use of traffic analysis (59). This indicates a blend of static and dynamic analysis techniques to uncover software logic flaws and privacy issues.

For in-vehicle components, vulnerabilities in IVI require extensive use of reverse engineering (146) and code audit (146), along with firmware extraction (135) and vulnerability reproduction (39). These skills point to the need for both static inspection and dynamic testing of services and interfaces.
ECU vulnerability discovery typically involves reverse engineering, code audit, and firmware extraction (19 each), supported by brute force (12), port scanning (13), vulnerability reproduction, and traffic analysis. These requirements suggest the need for direct hardware access and expertise in analyzing embedded firmware.

For the network components, the analysis primarily relies on brute force (15), port scanning (15), and code audit (7). Researchers often use tools like CANoe to inspect communication protocols such as CAN, LIN, and FlexRay, attempting to forge control messages or send malformed packets to trigger system malfunctions.

Participants employed a diverse range of technical approaches depending on the type of vulnerability being discovered. Authorization vulnerabilities were the most frequently analyzed and involved the broadest spectrum of skills. The most commonly used techniques included penetration testing (94), reverse engineering (90), and code audit (90). This is expected, as such vulnerabilities often require comprehensive analysis of authentication mechanisms, access control models, and communication protocols.

Information leakage and injection attacks were also common and primarily relied on static analysis techniques such as reverse engineering and code audit, often combined with penetration testing. These findings suggest that such vulnerabilities typically stem from flaws in communication logic or input validation, requiring a deep understanding of application-level business processes.

DoS attacks involved a more diverse set of skills. In addition to traditional methods, they often required vulnerability reproduction (9) and fuzzing (8), highlighting the value of test-driven techniques in verifying system stability and robustness. For Web-specific vulnerabilities, penetration testing was the sole technique employed (26 occurrences), underscoring the emphasis on dynamic interaction with web interfaces and business logic rather than firmware or code-level weaknesses.

\vspace{2mm}
\insightbox{
\revised{\textbf{Finding 6:}
ICV vulnerability discovery spans both software and hardware layers, requiring researchers to possess end-to-end analysis capabilities and cross-disciplinary expertise. In general, different vulnerability locations tend to demand different skill sets, highlighting the need for comprehensive and modular training in automotive security.
}
}

\vspace{2mm}
\insightbox{
\revised{\textbf{Finding 7:}
Different types of vulnerabilities require distinct skill sets.
Overall, penetration testing, reverse engineering, and code audit emerged as core capabilities applicable across nearly all vulnerability types, demonstrating their high versatility.
Instead, techniques such as fuzzing, traffic analysis, and unpacking were more context-specific, reflecting the need to adapt skill sets based on the nature of the vulnerability and the attack surface.}}

\subsection{ICV characteristics}
As described in Section~\ref{sec: icv profile}, the participating ICVs were categorized along four dimensions: country/region, EPA size, SAE automation level, and power type.
This section analyzes the correlation between these categories and both vulnerability locations and types, aiming to reveal distribution patterns and identify common security weaknesses across different classes of vehicles. As shown in Table~\ref{tab: locationplus}, we mainly put the correlations to EPA size, SAE level, and power type. Due to the page limit, the analysis of vulnerabilities by country/region is provided on our website~\cite{nationaplus}.
\begin{table}[t]
\centering
\footnotesize
\setlength{\tabcolsep}{1.0pt}
\caption{\revised{Mapping results of the ICV class taxonomy.}}
\begin{tabular}{c|c|ccc|ccc|ccc}
\toprule
\multicolumn{2}{c|}{\multirow{2}{*}{Location \& Type}} & \multicolumn{3}{c|}{EPA size} & \multicolumn{3}{c|}{SAE level} & \multicolumn{3}{c}{Power type}  \\
\cmidrule{3-11}
\multicolumn{2}{l|}{} & SUV & Sedan & MPV & L1& L2& L3& Electric &Hybird&Petrol\\
\midrule
\multirow{8}{*}{Location}& Cloud platform  & 63 & 38 & 4 & 6 & 95 & 4 &44&37&24  \\	
&IVI & 111 & 82 & 8 & 10 & 183 & 8 &74&93&34 \\
&APP&17 &41 & - &- & 46 & 12 &49&1&8 \\
&ECU&27 & 13 & - & 8 & 28 & 4 & 26 & 6 & 8 \\
&Network & 16 & 15 & - & 2 & 25 & 4 & 22 & 6 & 3  \\
&T-Box& 17 & 8& - &- & 25& -& 9 & 6 &10 \\
&Radio& 2 & 13 & - & - &15 & -&8&5&2  \\
&Charging pile &-& -&-&-&-&-&-&-&-\\ 
\midrule
\multirow{9}{*}{Type}&Authorization& 98 & 75 & 4& 8 & 161 & 8& 95 & 58 & 24 \\
&Information leakage & 37 & 35 & 3 & 9 & 57 & 9 & 45 & 9 & 21 \\
&Injection & 22 & 20 & 1 & 2& 39 & 2& 13 & 26 & 4\\
&OS& 40 & 13 & 1 & 2 & 50 & 2 & 19 & 26 & 9\\
&DoS & 17 & 26 & - & 2 & 35 & 6 & 25 & 9 & 9\\
&File operation& 16 & 3 & 3& 2& 19& 1& 3 &11& 8\\
&Interception& 12 & 24 & - & 1 & 32 & 3 & 21 & 9 & 6\\
&Web-specific& 7 & 4 & -&-& 10& 1& 2 & 6 & 3\\
&Memory& 4&10& - &-& 14&-& 9 & - & 5\\
\midrule
\multicolumn{2}{c|}{Sum}&253& 210& 12 &26& 417 & 32 &232&154&89 \\
\multicolumn{2}{c|}{Car count}  & 20 & 24 & 4 &6&36 &  6&22&16&10\\
\multicolumn{2}{c|}{\#Vulnerabilities/Vehicle}&12.7& 8.8& 3.0&4.3& 11.6 & 5.3 &10.5&9.6&8.9 \\

\bottomrule
\end{tabular}
\label{tab: locationplus}
\vspace{-4mm}
\end{table}

\textbf{EPA size mapping.}
The average number of vulnerabilities differs significantly across vehicle types with different EPA size classifications.
SUV models have the highest average number of vulnerabilities (12.7 per vehicle), far exceeding sedans (8.8 per vehicle) and MPVs (3.0 per vehicle).
As a highly competitive mainstream market segment, SUVs typically integrate more advanced connected infotainment and driver assistance features, which lead to a larger attack surface and greater potential risk~\cite{9348647}.
In terms of vulnerability types, authorization issues stand out as the most prominent security weakness across all EPA size categories.
They rank first in number for SUVs, sedans, and MPVs alike. This indicates that access control and permission management are pervasive and systemic security shortcomings, regardless of vehicle size classification.

\textbf{SAE level mapping.}
From the perspective of SAE levels, L2 vehicles have a significantly higher average number of vulnerabilities (11.6) compared to L1 (4.3) and L3 (5.3) vehicles.
Compared to L3 systems, L2 systems still rely heavily on interfaces (such as voice control), making IVI functions more complex and interfaces more abundant, thereby expanding the attack surface.
This suggests that the widely adopted L2 level system, due to their functional complexity and deep integration with the vehicle, are currently at a peak stage of vulnerability risk.

\textbf{Power type mapping.}
In terms of power types, there is a clear positive correlation between the average number of vulnerabilities and the degree of vehicle electrification and intelligence. Electric vehicles have the highest average number of vulnerabilities (10.5), followed by hybrid vehicles (9.6), while traditional petrol vehicles have the lowest (8.9). This demonstrates that as vehicles integrate more electronic control units, connected services, and remote control features, their overall security risks increase accordingly.

\vspace{2mm}
\insightbox{
\revised{\textbf{Finding 8:}
The number of vulnerabilities in a vehicle is related to its intelligence and functional complexity, which is particularly evident in SUVs, L2-level vehicles, and electric vehicles.}}}

\vspace{-2mm}
\section{Discussion}
\revised{This section discusses two aspects: implication for stakeholders and threats to validity.}
\vspace{-3mm}
\revised{
\subsection{Implications for stakeholders in ICVs}
Our empirical analysis reveals that the primary challenge in the ICV industry is not the inability to defend against advanced attacks, but in the systemic neglect of basic security best practices.
A large number of vulnerabilities stem from basic oversights, such as 75 instances of weak or plaintext passwords, 112 cases of information leakage, and 36 intercepting attacks (together accounting for 34.36\% of all vulnerabilities), as well as unprotected debugging interfaces and unnecessary open ports (e.g., SSH, Telnet).
Even more troubling, some known CVEs patched as early as 2017 remain unaddressed, exposing critical supply chain risks caused by the integration of outdated components for cost-saving purposes.
Therefore, improving ICV security hinges on the rigorous implementation of Secure Software Development Lifecycle principles, particularly in the areas of basic security hygiene and supply chain risk management.
Improving the security of ICVs requires coordinated efforts from all stakeholders.
To this end, manufacturers should strengthen basic security protections and third-party component management, while ensuring secure data transmission through encryption.
Researchers are encouraged to build ICV-specific vulnerability databases and develop efficient automated detection tools.
Regulatory bodies should enforce mandatory standards to unify baseline security requirements across the industry and enhance overall system resilience.
}

\vspace{-11mm}
\revised{\subsection{Threats to validity}}
\revised{While our study presents a large-scale empirical analysis of real-world ICV vulnerabilities, several factors may affect the reproducibility, replicability, and generalizability of our findings.

\textbf{Vehicle confidentiality.}
Due to confidentiality agreements and ethical considerations, we are unable to disclose detailed information about the participating vehicle brands, models, or firmware versions. This limits the ability to exactly reproduce or replicate our results. Additionally, our findings may not fully generalize to vehicles outside of our study. To mitigate this, we selected a diverse set of 48 commercially available ICVs across various architectures and manufacturers—an effort that required substantial cost and resources. We also provide high-level profiling of these vehicles, including SAE automation level, power type, and EPA vehicle size classification, to enhance contextual understanding.

\textbf{Participant expertise.}
Variations in participant skill sets could also influence the types of vulnerabilities discovered. Different teams may yield different results depending on their domain expertise, making it difficult for others to reconstruct teams with equivalent capabilities. To mitigate this, we provided participant skills based on submitted vulnerability reports and analyzed their association with different vulnerability types and locations.

\textbf{CTF competition constraints.}
The time constraints and incentive structures inherent in CTF-style competitions differ from those in long-term red-team engagements or real-world attack scenarios. As such, the findings in this study are most representative of controlled, time-bounded environments. Their applicability to persistent threat models or production systems should be interpreted with caution. To address this, we analyzed the vulnerability discovery process in detail, provided taxonomy coverage statistics, and released our framework to allow future CTF results to iteratively refine and extend our taxonomies.

\textbf{Taxonomy construction bias.}
Our unified taxonomy was constructed based on 13 carefully selected survey papers. While we aimed to include recent, comprehensive, and highly relevant works, the selection process could introduce bias and affect the taxonomy’s initial coverage. Nonetheless, the selected papers cover broad literature and provide a solid foundation for consolidation.

\textbf{Annotation subjectivity.}
To support reproducibility, we released a dataset of 649 validated vulnerabilities along with a multi-dimensional classification schema. However, we could not release technical artifacts such as proof-of-concept code due to ethical and legal restrictions. Additionally, the labeling process involved expert judgment, which may introduce subjectivity and potential data incompleteness. To mitigate this, all classification and validation tasks were conducted by four highly experienced experts in ICV security, following multiple rounds of discussion to reach consensus.
}
\vspace{-1mm}

\vspace{-3mm}
\section{Conclusions}

This study presents a systematic analysis of vulnerabilities in Intelligent Connected Vehicles. Through comprehensive testing efforts, we constructed a dataset of 649 verified vulnerabilities, spanning 48 different ICV models. \revised{Based on prior literature, we extracted and unified vulnerability taxonomies along two key dimensions: location and type. We then mapped the collected vulnerabilities to these categories to assess how well existing taxonomies reflect real-world threats and to identify newly emerging categories uncovered through empirical data. In addition, we found significant correlations between vulnerability characteristics and vehicle types, further highlighting the importance of implementing modular and differentiated protection strategies.}




\bibliographystyle{ACM-Reference-Format}
\bibliography{refs}
\end{sloppypar}
\end{document}